\documentclass[usenatbib]{mn2e}
\usepackage{aas_macros,psfig,amsmath}

\begin{document}
\topmargin-1cm

\def\bi#1{\hbox{\boldmath{$#1$}}}

\newcommand{\be}{\begin{equation}}
\newcommand{\ee}{\end{equation}}
\newcommand{\beqa}{\begin{eqnarray}}
\newcommand{\eeqa}{\end{eqnarray}}

\newcommand{\lexp}{\mathop{\langle}}
\newcommand{\rexp}{\mathop{\rangle}}
\newcommand{\rexpc}{\mathop{\rangle_c}}

\newcommand{\etal}{et al.\ }
\newcommand{\kms}{\, {\rm km\, s}^{-1}}
\newcommand{\ikms}{(\kms)^{-1}}
\newcommand{\mpc}{\, {\rm Mpc}}
\newcommand{\kpc}{\, {\rm kpc}}
\newcommand{\hmpc}{\, h^{-1} \mpc}
\newcommand{\ihmpc}{(\hmpc)^{-1}}
\newcommand{\hkpc}{\, h^{-1} \kpc}
\newcommand{\lya}{Ly$\alpha$}
\newcommand{\lyaf}{Ly$\alpha$ forest}
\newcommand{\bF}{\bar{F}}
\newcommand{\xrei}{x_{rei}}
\newcommand{\lr}{\lambda_{{\rm rest}}}
\newcommand{\hi}{\mbox{H\,{\scriptsize I}\ }}
\newcommand{\heii}{\mbox{He\,{\scriptsize II}\ }}
\newcommand{\vdF}{{\mathbf \delta_F}}
\newcommand{\PF}{P_F(k,z)}
\newcommand{\PL}{P_L(k_p,z_p)}
\newcommand{\gprime}{g^\prime}
\newcommand{\sprime}{s^\prime}
\newcommand{\tgamma}{\tilde{\gamma}}
\newcommand{\gmo}{\gamma-1}
\newcommand{\Tp}{T_{1.4}}
\newcommand{\DL}{\Delta_L^2(k_p,z_p)}
\newcommand{\neff}{n_{\rm eff}(k_p,z_p)}
\newcommand{\aleff}{\alpha_{\rm eff}(k_p,z_p)}
\def\pvm#1{[PM: {\it #1}] }
\def\US#1{[US: {\it #1}] }

\def\bi#1{\hbox{\boldmath{$#1$}}}

\title{Physical effects on the \lyaf\ flux power spectrum: damping wings, 
ionizing radiation fluctuations, and galactic winds} 

\author[Patrick McDonald, Uro\v s Seljak, Renyue Cen, Paul Bode, and Jeremiah P. Ostriker]
{Patrick McDonald$^1$\thanks{Electronic address:
    {\tt pm@princeton.edu}}, Uro\v s Seljak$^1$,  
Renyue Cen$^2$, Paul Bode$^2$ and Jeremiah P. Ostriker$^{2}$
\\$^1$Department of Physics, Jadwin Hall, Princeton University, Princeton, NJ 08544, USA
\\$^2$Princeton University Observatory, Princeton University, Princeton NJ
      08544, USA
}

\maketitle

\begin{abstract}

We explore several physical effects on the power spectrum of the \lyaf\
transmitted flux.  The effects we investigate here are usually not part 
of hydrodynamic simulations and so need to be estimated separately.  The most 
important effect is that of high column density absorbers with damping wings, 
which add power on large scales.  We compute their effect using the 
observational constraints on their abundance as a function of column density.  
Ignoring their effect leads to an underestimation of the slope of the linear 
theory power spectrum.  The second effect we investigate is that of 
fluctuations in the ionizing radiation field.  For this purpose we use a very 
large high resolution N-body simulation, which allows us to simulate both the 
fluctuations in the ionizing radiation and the small scale \lyaf\ within the same 
simulation.  We find an enhancement of power on large scales for quasars and a 
suppression for galaxies.  The strength of the effect rapidly increases with 
increasing redshift, allowing it to be uniquely identified in cases where it is 
significant.  We develop templates which can be used to search for this effect 
as a function of quasar lifetime, quasar luminosity function, and attenuation 
length.  Finally, we explore the effects of galactic winds using hydrodynamic 
simulations.  We find the wind effects on the \lyaf\ power spectrum to be be 
degenerate with parameters related to the temperature of the gas that are 
already marginalized over in cosmological fits.  While more work is needed to 
conclusively exclude all possible systematic errors, our results suggest that, 
in the context of data analysis procedures where parameters of the \lyaf\ model 
are properly marginalized over, the flux power spectrum is a reliable tracer of 
cosmological information.

\end{abstract}

\section{Introduction}

One of the primary goals of cosmology is the determination of the 
origin of structure in the universe.  Among the available probes today 
one of the most promising 
is the amplitude and shape of the power spectrum of the primordial
fluctuations.
It can be constrained both by the CMB and other probes
of large scale structure. Combining different data sets that cover a large
range in scale is particularly powerful. Of the current cosmological
probes, the Ly-$\alpha$ forest -- the absorption observed in quasar spectra
by neutral hydrogen in the intergalactic medium (hereafter IGM) --
has the potential to give the most precise information on small scales.
It probes fluctuations around megaparsec scales at redshifts between
2-4, so nonlinear evolution, while not negligible, has not erased
all of the primordial information.
The statistic of choice when analyzing the \lyaf\ is the 
power spectrum of the transmitted flux fraction, $\PF$ 
\citep{1998ApJ...495...44C}.
While several groups have extracted this quantity from the 
data \citep{2002ApJ...581...20C,2000ApJ...543....1M,2003astro.ph..8103K},
a recent analysis of 3300 quasar spectra from the Sloan Digital Sky 
Survey (SDSS) 
has improved the statistical errors by an order of magnitude 
\citep{2004astro.ph..5013M}. This significant increase in statistical 
power must be accompanied by a 
corresponding improvement in the theoretical calculations 
if the new data are 
to be exploited to the fullest extent. The goal of this paper is to make 
a step in this direction. 
We focus on three topics: 
high density systems (especially the effect of damping wings), 
fluctuations in the ionizing background, and effects of
galactic superwinds. 

In the standard picture of the Ly-$\alpha$ forest
the gas in the IGM is in ionization equilibrium.  The
rate of ionization by the UV background balances the rate of
recombination of protons and electrons.
The recombination rate depends on the temperature of the gas, which
is a function of the gas density.
The temperature-density relation can be parameterized by an amplitude,
$T_0$, and a slope $\gamma-1=d\ln T/d\ln \rho$.
The uncertainties in the mean intensity of the UV background, the mean
baryon density, and other
parameters that set the normalization of the relation between optical
depth and density can be combined into one parameter: the mean transmitted
flux, $\bar{F}(z)$.
An additional nuisance parameter is the filtering
length $k_F$ \citep{1998MNRAS.296...44G}, which is related to the 
thermal history of the IGM.  
The parameters of the gas model, $T_0$, $\gamma-1$, $\bar{F}$, and $k_F$,
are marginalized over when computing constraints on cosmology
\citep{2004astro.ph..7377M}.
Thus any additional physical effects must be different from the
combined effect of these 4 parameters if they are to be relevant for the 
cosmological analysis. Our analysis differs from
previous analyses in that we include this factor when assessing the 
importance of a given physical effect. 

The first effect we investigate is that of high density absorbers 
on the flux power spectrum.
In the bulk of the intergalactic medium (IGM) at high redshift,
neutral gas is in ionization equilibrium, with the proton-electron 
recombination rate balancing ionization of neutral hydrogen by a
fairly uniform ionizing background.
Systems with neutral column density 
$N(\hi)>{\rm few}\times 10^{17}{\rm~atoms~cm^{-2}}$ become 
self-shielded, i.e.,
the exterior of the system absorbs all incoming ionizing radiation so the
interior no longer sees the background, becoming mostly neutral.
These absorbers are called Lyman-limit systems (LLS).
LLSs should be relatively small 
\citep{2001ApJ...559..507S}, and for that or other reasons, such as 
too small simulation boxes, are 
not generally well reproduced in hydrodynamic simulations 
\citep{1996ApJ...471..582M,
2001ApJ...559..131G,2003ApJ...598..741C,2004MNRAS.348..421N}.
As the column density of the systems increases toward the traditional
definition of a damped-\lya\ absorber (DLA), 
$N(\hi) > 2\times 10^{20}{\rm~atoms~cm^{-2}}$ \citep{1986ApJS...61..249W,
1986ApJ...310..583S}, damping wings
come to dominate the equivalent width of the systems, enhancing their
impact on spectra.  In addition to the problem of simulations 
not necessarily 
reproducing the number of systems, damping wings are not usually
included in the simulated spectra used to predict $\PF$ at all, 
because the widest of them can extend to the full width of
a typical simulation box.

The effect of damping wings differs from the effects of UV background
fluctuations and galactic winds, discussed next, in that they are certainly
present in the data, and the systems are even more or less directly 
observable, although the detailed column density distribution below 
$2\times 10^{20}{\rm~atoms~cm^{-2}}$ is more difficult to resolve
observationally and has not been the subject of much investigation.
The possible importance of DLAs was recognized by 
\cite{1999ApJ...520....1C}, who investigated their effect by measuring
$dP_F/dk$ with and without the identified systems in their spectra, 
finding a change
that was generally not larger than their statistical error bars.
\cite{2004MNRAS.349L..33V} emphasized the non-negligible contribution
of systems with $N(\hi) > 10^{15}{\rm~atoms~cm^{-2}}$ to the flux 
power.  
In this paper we investigate the effect of
high density systems in general in simulations, and particularly the 
importance of the damping wings of absorbers with 
$N(\hi) < 2\times 10^{20}{\rm~atoms~cm^{-2}}$.

The second effect we analyze in this paper are the fluctuations in 
the ionizing background.  
The intensity of the ionizing background determines the density of neutral 
hydrogen and if the intensity is spatially varying this will lead to spatial 
variations in neutral hydrogen density.
These intensity fluctuations in the ionizing background and their 
effect on the \lyaf\
have been discussed in the past 
\citep{1992MNRAS.258...36Z,1993ApJ...415..524F,1999ApJ...520....1C,
2002MNRAS.334..107G}.  The two most recent works 
investigated it in some detail focusing on \lyaf\ statistics such as 
the flux 
power spectrum \citep{2004MNRAS.350.1107M,2003astro.ph.10890C}. 
\cite{2004MNRAS.350.1107M} use relatively small dark matter PM simulations 
to investigate 
the effect, focusing on high redshifts ($z>4$) where attenuation lengths 
are short and the small simulation box sizes used 
are adequate. They assume quasar hosts 
are uncorrelated with the \lyaf\ and as a consequence they find the effect 
to enhance the fluctuations on large scales. In contrast, 
\cite{2003astro.ph.10890C} used
large $P^3M$ dark matter simulations in combination with small scale 
hydro-dynamic simulations to investigate the effect at $z=3$. In this 
model quasar hosts are placed in high density regions and are correlated
with the \lyaf. 
Since there is a higher intensity ionizing background in the 
regions where there is also more neutral hydrogen, a cancellation
occurs and fluctuations in the \lyaf\ are suppressed on large scales. 
\cite{2003astro.ph.10890C} introduced several additional improvements 
in the modeling 
of the effect, such as inclusion of light cone effects due to 
quasar finite lifetime, photon shadowing by high density 
neutral hydrogen along the line of sight, which was investigated using a 
photon ray tracing code, and an investigation of the effects of beaming. 
While 
the photon shadowing and beaming appear
to be of little relevance for the \lyaf\ flux power spectrum statistics,
the light cone effects increase the effect by up to a factor of two. 
In this paper we will therefore ignore shadowing and beaming, 
using the uniform isotropic attenuation approximation, 
but we will include the lightcone effects. 

The main numerical issue when investigating the effect of ionizing 
background fluctuations on the \lyaf\ statistics is the required 
dynamic range of simulations. The ionizing background attenuation length 
in comoving units
ranges between 50$\hmpc$ and 500$\hmpc$ in comoving units, 
so to properly estimate the fluctuations 
box sizes of order several hundred megaparsecs are required. On the 
other hand, to properly simulate the \lyaf\ one needs resolution below
100$h^{-1}$kpc to resolve the Jeans scale. 
Combining the two requirements leads to a dynamic range not 
available presently in hydrodynamic simulations. 
In \cite{2003astro.ph.10890C} this problem was approached
using a hybrid method where large scales
are simulated with a dark matter only simulation and a hydrodynamic 
simulation is used to simulate small scales, by randomly choosing a patch that 
matches the $P^3M$ simulation in the value of the density field 
smoothed at a larger
scale. However, this approach does not preserve all of the correlations 
present in a single full resolution simulation and it is not clear 
if the results can be used for quantitative estimates. 

In this paper we use pure dark matter simulations to simulate both 
ionizing background fluctuations and the \lyaf. We perform the 
analysis as a function of redshift over the range $2<z<4.6$.  
Dark matter simulations can reproduce qualitatively 
full hydrodynamic simulations 
\citep{1998MNRAS.296...44G,2001MNRAS.324..141M}. 
Our approach is similar to \cite{2004MNRAS.350.1107M}, 
where it was shown, using a series of resolution studies, that to assess 
the relative effects between simulations with and without fluctuations 
it suffices to have a $128^3$ simulation in a 30$\hmpc$ box, even if the small 
scale fluctuations in the \lyaf\ are under-resolved in an absolute sense. 
In this paper we use a $1024^3$
TPM simulation in a 320$\hmpc$ box \citep{2003ApJS..145....1B}, 
whose mass resolution and force 
resolution in low density regions 
is similar to a $128^3$ PM simulation in a 30$\hmpc$ box, 
while its force resolution in high density regions 
is significantly better. The size of the simulation is several 
attenuation lengths at $z=3$ and we use periodic boundary conditions to extend 
the calculations of photon propagation from quasars to even larger distances. 
Thus we expect this simulation to be sufficiently accurate for our
purposes both on large and small scales. 
An additional advantage of its high mass and force resolution is 
that it resolves halos with masses above $10^{11}h^{-1}M_{\sun}$. 
This allows us to generate a halo catalog using
a friends-of-friends halo finder, which we
used as a catalog of quasar hosts. 

The third physical effect on the \lyaf\ that we address in this paper 
is that of galactic superwinds (GSW). There is clear evidence that 
powerful winds emerging out of galaxies 
are present both at low redshifts \citep{2003RMxAC..17...47H} and at high 
redshifts \citep{2001ApJ...554..981P}, but the extent to which these 
affect the intergalactic 
medium (IGM) at large separations from the host galaxy is poorly known. 
Current studies of \lyaf\ absorption in background quasars  
at small angular separations from a Lyman-break galaxy (LBG) suggest that 
winds (or something like them) 
may be affecting IGM to an appreciable distance 
\citep{2003ApJ...584...45A}, but the 
statistics are poor and there could be selection effects involved. 
However, there is also ample evidence of metals in the intermediate to 
low density IGM
\citep{1996AJ....112..335S,2003ApJ...596..768S,2004A&A...419..811A}, 
which suggests that GSW must be operating in transporting 
metals from galaxies to the lower density IGM. 

There have been several recent simulations investigating the effect 
of GSW on the \lyaf\ 
\citep{2002ApJ...580..634C,2002ApJ...580...42M,2003MNRAS.343L..41B,
2003ApJ...594...75K,2004MNRAS.350L..21M,2003astro.ph.11209D}. 
Many of these focused on the
correlations with LBGs, attempting to reproduce the observational 
results of \cite{2003ApJ...584...45A}. While they are 
generally successful in  reproducing the 
observations at larger distances, they fail in the 
inner $\sim h^{-1}$Mpc, where the observations
suggest that there is frequently little or no absorption, 
while simulations would suggest that the 
absorption is significantly larger because of the increased density of 
neutral hydrogen close to the galaxy. 
In \cite{2002ApJ...580..634C,2003astro.ph.11209D} 
the effects of winds on the \lyaf\ flux power spectrum have also been 
investigated. 
For \cite{2002ApJ...580..634C} the effects are small, but not negligible 
at the level 
required by new \lyaf\ measurements \citep{2004astro.ph..5013M}, while in 
\cite{2003astro.ph.11209D} the effects appear to be 
completely negligible. 
To explain the observations the winds were
assumed to blow out a bubble around LBGs, 
where the bubble radius was put in by hand.  
Some of these comparisons were based simply on comparing wind and 
no wind simulations to each other, 
while winds produce other effects such as heating 
of the IGM. In this case it is not clear if the effects 
produced by the winds are unique or degenerate 
with changes in the mean absorption level,
temperature-density relation, or filtering length, all of which we 
marginalize over in the standard analysis anyways. 
In this paper we will address
these issues by looking for residual effects after these effects are 
accounted for. 
We will use wind simulations from \cite{2004astro.ph..7143C}, 
which have been calibrated to the observed wind velocities, 
but more extreme cases are also explored. 

\section{High Density Absorbers}

In this section we investigate the effect of high density
absorbers on $\PF$.
The investigations by 
\cite{1999ApJ...520....1C} and \cite{2004MNRAS.349L..33V} were
not performed at the level of detail we need
to compare to the $\PF$ measurement of \cite{2004astro.ph..5013M}.  
The \cite{2004astro.ph..5013M}
measurement and the accompanying simulation analysis by \cite{2004astro.ph..7377M} 
spans the redshift
range $2.2\sim< z\sim<4.2$ in bins spaced by $\Delta z=0.2$, and probes 
scales $0.0013~{\rm s/km} < k < 0.02~{\rm s/km}$ in bins spaced by 
$\Delta \log_{10}(k)=0.1$.  The precision of the 132 resulting $\PF$ 
points is usually better than 10\% and often as good as 3\%.  Our goal
in this section, once we find an effect significant enough to require us
to account for it when fitting to the data, is to create templates for
the dependence of the effect on $k$ and $z$, with a level of
accuracy sufficient to match the present data.

Within the context of the model for the IGM that we are 
assuming, as represented in the hydrodynamic simulations,
systems with $N(\hi)<1.6\times 10^{17}{\rm~atoms~cm^{-2}}$ 
should not present any problem for us, so we will focus
on higher column density systems.  The 
$N(\hi)<1.6\times 10^{17}{\rm~atoms~cm^{-2}}$ systems are close
enough to resolved \citep{2001ApJ...559..507S} in simulations
that if they are important to the 
power spectrum, in a way that is not yet fully resolved, the 
power spectrum would change as the resolution is increased in
standard tests such as those performed in \cite{2004astro.ph..7377M}
(i.e., there is no reason not to expect a smooth change in 
the power spectrum with increasing resolution).
Systems with $N(\hi)>1.6\times 10^{17}{\rm~atoms~cm^{-2}}$
seem more likely to be a problem because the onset of
self-shielding causes a rather sudden increase in 
optical depth with density, the importance of which might
not be detected by resolution tests (this is not obvious,
but is conceivable).

The section is arranged as follows:
In \S \ref{obscold}, we explain 
the observationally
determined column density distribution that we use 
in \S \ref{simhighdense} and \ref{dampingwings}.
Then in \S \ref{simhighdense} we look at high density absorbers
in the hydrodynamic 
simulations used to predict the power spectrum in 
\cite{2004astro.ph..7377M}, without considering damping wings.  Finally,
in \S \ref{dampingwings}, we 
estimate the effect of the damping wings of the highest density
systems on $\PF$.  

\subsection{Estimate of the $\log(N)>17.2$ 
Column Density Distribution from 
Observations \label{obscold}}

Subsections \ref{simhighdense} and \ref{dampingwings} rely on an
estimate of the column density distribution, $f(N)$, for all
$N(\hi)>1.6\times 10^{17} {\rm~atoms~cm^{-2}}$.  Traditionally,
the full distribution has only been measured for
$N(\hi)>2\times 10^{20} {\rm~atoms~cm^{-2}}$, with the LLSs
grouped together into a total number density $n(z)$, because
the column densities of these systems are hard to measure.
In the course of our investigation of damping wings,
we determined that the systems with $\log(N)<20.3$ 
were significant, so we decided to re-estimate the
full column density distribution from the data.

The number
of systems as a function of column density, $f(N)$,
is expected to have a non-trivial form in the 
LLS-DLA regime, because of self-shielding.  
If one imagines increasing the  
density of a reference system, the implied neutral 
density increases gradually in ionization
equilibrium with the background radiation
until the system reaches $\log(N)\sim 17.2$, 
when self-shielding reduces the effective ionizing 
background and the neutral density suddenly increases
much more rapidly until the system becomes almost
completely neutral.  \cite{2002ApJ...568L..71Z} give
predictions for the resulting column density distribution,
based on a spherical isothermal halo model.  The model 
has two free parameters:  the overall normalization 
of $f(N)$ (proportional to the number density of halos), 
and a parameter that we will nominally call the mass of the
halos, although this interpretation should not be taken
literally as the value of the parameter depends on other
things like the strength of the ionizing background
\citep{2002ApJ...568L..71Z}.    The \cite{2002ApJ...568L..71Z}
model is a toy model, but it seems reasonable to assume
that the class of $f(N)$ shapes that it describes 
contains a shape close enough to the truth for our
purposes.

The two parameters, amplitude and halo mass, are 
not predicted by the model so we will determine them
by a fit to the data.
To describe the column density distribution including
dependence on redshift, we perform
a maximum likelihood fit using the formula 
\begin{equation}
f(N,z)=A~x_z^{\gamma_A}
f_{\it ZM}(N,M_\star~x_z^{\gamma_M})~,
\end{equation}
where $x_z=(1+z)/(1+z_\star)$, and $f_{\it ZM}$ is the
functional form described by \cite{2002ApJ...568L..71Z}, 
i.e., we allow 
the overall normalization to evolve as a power law 
in $1+z$, and allow the mass of the typical halo to
evolve similarly.  
In practice, 
we include the $M$ dependence by simply interpolating
between the four example curves in \cite{2002ApJ...568L..71Z}'s 
Figure 2 
(using linear interpolation in $\ln(f)$ and $\ln(N)$),
which were provided in numerical form by Z. Zheng.  

We take the data in the DLA regime from \cite{2004astro.ph..3391P}.
For reference, \cite{2004astro.ph..3391P} estimate the number density 
of DLAs per unit redshift, 
finding $n_{\rm DLA}(z)=0.189 \pm 0.026$
for a redshift bin with $2.0 < z < 2.5$, 
$0.215\pm 0.032$ for $z=2.5-3.0$, $0.271\pm 0.049$ for
$z=3.0-3.5$, $0.366 \pm 0.073$ for $z=3.5-4.0$, and
$0.401 \pm 0.121$ for $z=4-5$.  
We do not use these numbers, or the 
\cite{2004astro.ph..3391P} column density distribution
directly, because they would not give an optimal constraint
on our model parameters.  Instead, our maximum likelihood 
analysis
uses the column densities of individual absorbers and 
the segments of quasar spectra searched, 
considering only $z>2.0$ (these data were provided in 
machine readable form by J. Prochaska).

Recently, \cite{2003MNRAS.345..480P} made a first measurement 
of the abundance of systems with 
$1.0\times 10^{19} {\rm~atoms~cm^{-2}} < N(\hi)
< 2.0\times 10^{20} {\rm~atoms~cm^{-2}}$, finding
$n_{19-20.3}(z<3.5)=0.49\pm ?$ and 
$n_{19-20.3}(z>3.5)=0.97\pm ?$ (there were a total of 7 and 3 
systems, respectively, from which one can estimate the unsupplied
error bars).  We use their data quasar-by-quasar, reading
from their Table 1, in the
same way that we use the DLA data.

Finally, we add a constraint on the integral of $f(N)$
over the LLS column density range, $n_{\rm LLS}(z)$
from \cite{2003MNRAS.346.1103P}.  They measured 
$n_{\rm LLS}(z)=n_0 (1+z)^\gamma$ 
LLS per unit redshift, with $n_{\rm LLS}=0.07^{+0.13}_{-0.04}$
and $\gamma=2.45^{+0.75}_{-0.65}$, for LLS with 
$2.4 \sim < z \sim < 4.5$; however, note that the error 
on $n_0$ is useless because the correlation with
$\gamma$ is not given -- the error on $n(z)$ at the 
center of weight of the data will be smaller than that
implied by the errors on $n_0$.
\cite{2003MNRAS.346.1103P}'s sample included 67 LLSs, 
implying that the error on the number density should be
about 12\%.

The results of the maximum likelihood fit 
are $A=25.1\pm 3.6$, $\gamma_A=1.80\pm 0.80$,
$\log(M/M_{\sun})=9.36\pm 0.18$, and $\gamma_M=1.4 \pm 2.4$,
for $z_\star=3.0$.
We give the best fit numbers only so that the column density 
distribution we use will be reproducible.  Our 
application only requires that something resembling 
the true column 
density distribution falls within the class of 
curves explored by varying $M$, which seems likely -- 
the interpretation of $M$ as a halo mass is much more
complicated, as discussed by \cite{2002ApJ...568L..71Z} 
(among other problems,
these curves were computed for an Einstein-de Sitter
Universe).  Note also that the errors are generally 
correlated. 
Performing an analysis of this type aimed at better understanding
LLSs and DLAs through their column density distribution would be 
an interesting topic for future work.

\subsection{Self-shielding and High Density Absorbers in 
Simulations \label{simhighdense}}

We start by examining the case where we ignore damping 
wings, because the wings require a different 
treatment using mock spectra that can be much longer
than the width of the simulations.  
We will see that this is ultimately an
academic exercise,
because the damping wings turn out to be the most 
important aspect of high density systems, but it is 
informative nonetheless.

The hydrodynamic simulations used in \cite{2004astro.ph..7377M}, 
which we will focus
on in this subsection, are described in more detail in 
\cite{2003ApJ...598..741C}.  We simulate a flat
$\Lambda$CDM cosmological model with $\Omega_m=0.3$, 
$\Omega_b=0.04$, and $h=0.7$, using an $L=10\hmpc$
box, with the gas properties tracked in $N=256^3$ 
fixed grid cells and the dark matter traced by $256^3$
particles.  
While \cite{2003ApJ...598..741C} reproduces the DLA abundance 
after allowing for dust obscuration, 
very high density systems in general are not necessarily 
well reproduced
by ours or any other hydrodynamic simulations
\citep{2003ApJ...598..741C,1996ApJ...471..582M,
2001ApJ...559..131G,2004MNRAS.348..421N}. 
Even if they were, 
we could not compute the effect of strong damping 
wings on our power spectrum accurately because of the 
limited width of our simulation boxes.

We start by estimating the effect of self-shielding in our
hydrodynamic simulations.  The simulations were performed
with a rough self-shielding approximation:  the background
radiation for each cell was attenuated by the column density
of that cell.  Assuming ionization equilibrium, we can 
recompute the neutral densities with and without this 
shielding (in the shielded case three iterations of the
density calculation, starting with no shielding, are 
required for convergence).  The effect of self-shielding
on the amount of neutral gas at $a=0.24$ is dramatic -- 
an increase
by a factor 70; however, in the absence of damping wings
the effect on $\bF$ is tiny -- 0.67561 is reduced 
to 0.67559.  The differences are the same size 
($\sim$0.00002) at $a=0.2$ and $0.32$, and the difference
in $\PF$ is never more than 0.1\% over the range
$0.0013~{\rm s/km} < k < 0.05~{\rm s/km}$.  This remarkable result
is not hard to understand:  in the absence of damping wings,
the equivalent width of an absorption line susceptible to
self shielding, i.e., with 
$N(\hi)>1.6\times 10^{17} {\rm~atoms~cm^{-2}}$, increases
by only $\sim$10\% per order of magnitude increase in $N(\hi)$,
and these lines are very rare to begin with.  This is an example
of a more general fact about the \lyaf\ -- it simply is not very
sensitive to the details of the high density regions of the 
Universe, except, 
as we will see, when damping wings become important.

Our simulations in fact produce too few LLS, even with self-shielding.  
To be sure that the missing systems do not significantly
affect our results, we tried simply arbitrarily increasing the column 
density of the most dense systems in the simulation so that we match
the observed number of LLS.  The affect is slightly larger than the
affect of self-shielding, but still tiny -- $\bF$ decreases by 0.0004,
and $\PF$ changes by less than 0.8\%.

In summary:  Without damping wings, LLS and denser systems are
completely irrelevant to the use of $\PF$ for cosmology.  
Self-shielding, despite changing the amount of neutral gas by
orders of magnitude, does nothing to the bulk of the IGM, and
there is no significant different between the mean absorption or 
$\PF$ predicted by our unmodified simulations and simulations 
where we have artificially inserted the observed number of LLS.

\subsection{Damping Wings \label{dampingwings}}

We now turn to the effect of damping wings.
Traditionally, LLS have been defined to be systems 
with $1.6\times 10^{17} {\rm~atoms~cm^{-2}} < N(\hi) 
< 2.0\times 10^{20} {\rm~atoms~cm^{-2}}$,
where $N(\hi)$ is the neutral hydrogen 
column density, while DLAs have 
$N(\hi) > 2.0\times 10^{20} {\rm~atoms~cm^{-2}}$.
This cutoff is somewhat arbitrary since systems with
$N(\hi)<2\times 10^{20}$ still show substantial 
damping wings.  For example, the equivalent widths 
of cold systems with $\log[N(\hi)]=$(17, 18, 19, 20, 21) 
are (54, 171, 539, 1705, 5392)$\kms$, while the 
widths at the same densities would be 
(164, 185, 204, 222, 237)$\kms$ 
if they were purely
thermally broadened at temperature $\sim 50000$ K.
We will explore the effect of absorption by these
systems on $\PF$ and $\bF$, using the observation
constraints on their column density from the next
subsection, combined with several models 
for their spatial distribution.

\begin{figure}
\centerline{\psfig{file=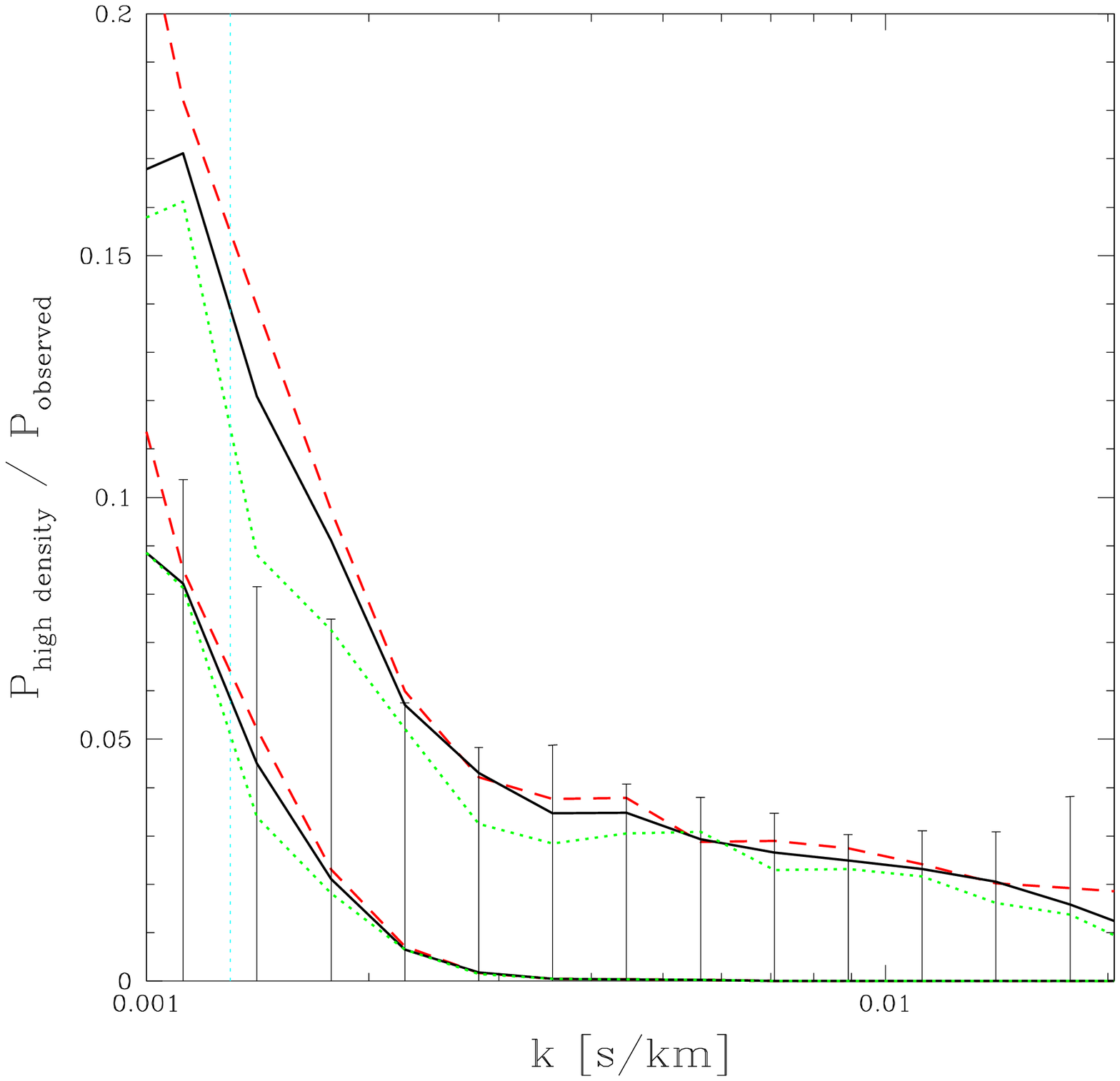,width=3.5in}} 
\caption{
$\PF$ from a random distribution of LLS and DLA (upper lines)
or just DLA (lower lines), each with the observed column density
distribution, relative to the observed \lyaf\ $\PF$ from 
\citep{2004astro.ph..5013M}.  
Red (dashed), black (solid), and green (dotted) lines show $z=2.2$,
3.2, and 4.2.  The error bars indicate the fractional error on the
observed $\PF$ at $z=3.2$ (the errors at $z=2.2$ are very similar,
while the errors at $z=4.2$ are $\sim 2$ times bigger).  The 
vertical dotted line shows the largest scale used for constraining
cosmology by \citep{2004astro.ph..7377M}.
}                       
\label{randomHDonly}
\end{figure}

The power spectrum of a completely
random distribution of systems is shown in Figure
\ref{randomHDonly}, along with the power from DLAs
only (the column density distribution used is 
described in \S \ref{obscold}).  
We see that the effect is quite significant, and 
that more than half of it (on relevant scales) 
comes from the $\log(N)<20.3$ systems.

\begin{figure}
\centerline{\psfig{file=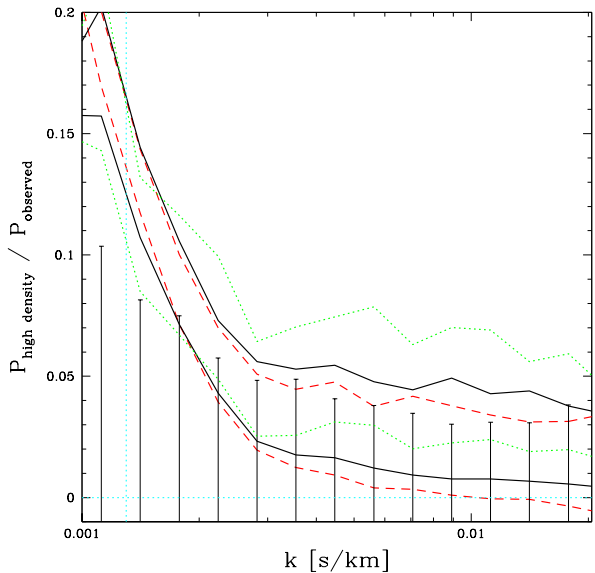,width=3.5in}}
\caption{
Change in $\PF$ when LLS and DLA with the observed 
column density distribution are inserted into mock
spectra of the \lyaf, relative to the observed \lyaf\ $\PF$.  
The upper curves show the case where the high density 
systems are inserted randomly (different from Fig. \ref{randomHDonly}
in that there was no background forest in that figure), 
while for the lower
curves the LLS and DLA were inserted at the highest density maxima 
in the mock \lyaf\ spectra.
The line types and error bars are as in Fig. \ref{randomHDonly}.
Note that a consistent systematic error that is $1\sigma$ for any 
single point is very significant, because we have many $z$ and $k$ 
bins to average over.
}
\label{mockHD}
\end{figure}

The region
of a simulated spectrum that is being replaced by the 
high density system may be relevant, 
as we saw in the case with no damping wings, 
We include this effect using the lognormal model
mock spectra described in \cite{2004astro.ph..5013M}.  
These are not perfectly realistic, but they have the
advantage over full numerical simulations that they 
can be of arbitrary length so the full extent of 
damping wings can be included.  Figure \ref{mockHD}
shows the power added to the mock spectra when
high density systems with the observed column density
distribution are added to the mock spectra, either at
random locations, or using a mapping of the highest
column density systems to the highest density maxima
in the lognormal field.
We see that the location of the systems does matter.
The change in high-$k$ power is removed when the 
systems are inserted at already high-density locations,
presumably because this power is related to relatively
low column density systems that are not significantly
broadened by damping wings and thus produce absorption
that is insensitive to column density, as seen above.
In fact, the difference between the added power in the
two cases is not
significant to cosmological fits \citep{2004astro.ph..7377M},
apparently because the fractional effect is is relatively 
flat (i.e., $k$-independent) 
and thus mostly degenerate with $\bF$ instead of the
primordial power spectrum.  The agreement in the
\cite{2004astro.ph..7377M} fits between these
two cases makes us confident that detailed, realistic 
simulations
of high density systems will not produce a significantly
different result.  The primary effect comes simply from
the self-correlation across the width of individual 
systems with damping wings,
which is directly computable from the observed column 
density distribution.  

How should we use these results to correct for their
influence on the determination of the linear theory
power spectrum, $P_L(k)$? The effect is essentially
additive, so the form 
$P_{\rm theory}(k,z)= P_{\rm sim}(k,z)+P_{\rm damp}(k,z)$
is most natural, where 
$P_{\rm theory}(k,z)$ is the final prediction to be
compared with observations, $P_{\rm sim}(k,z)$ is the
prediction by simulations without damping wings, and
$P_{\rm damp}(k,z)$ is the power shown in 
Figure \ref{mockHD}.  $P_{\rm damp}$ carries
some uncertainty of course.
As discussed in \S \ref{obscold},
the statistical errors on the number of LLSs and DLAs
are probably about 15\%, and the statistical error on
$P_{\rm damp}$ should be proportional to this.  
However, because most observations do not distinguish
column densities of LLSs, the important range of column 
densities just below the DLA cutoff is not directly 
constrained to the 15\% level.  For this reason, along
with general caution to allow for any errors in our
calculation, we suggest applying a 30\% error 
to the amplitude of $P_{\rm damp}(k,z)$ in 
fits, i.e., adding $A P_{\rm damp}(k,z)$ to the simulated
power, and $[(A-1)/0.3]^2$ to $\chi^2$.  An even more
cautious analysis might allow for redshift dependence of
$A$; however, there is not any reason to think a break from 
the \cite{2002ApJ...568L..71Z} column density distribution
template used in \S \ref{obscold} should be significantly
redshift dependent, and one might even argue that the 
constancy of the ratio of 
of $P_{\rm damp}(k,z)$ to $P_{\rm observed}(k,z)$ (Fig.
\ref{mockHD}) is unlikely to be a coincidence.
\cite{2004astro.ph..7377M} find a value of $A=1.01\pm0.24$, i.e.,
the power spectrum fit not only prefers our estimated
amplitude, but is beginning to constrain the amplitude
to better than our conservative 30\% external constraint.
One might think that a more careful study could reduce 
this error and thus reduce the errors on the final $P_L(k)$ 
measurement; however, \cite{2004astro.ph..7377M} shows that 
improving the error to 10\% does not noticeably improve
the $P_L(k)$ results (this may change as other parts of
the $P_L(k)$ analysis are improved).

\section{Fluctuations in the ionizing background}

The two most recent studies of fluctuations in 
the ionizing background and their effect on the flux 
power spectrum are \cite{2004MNRAS.350.1107M} and \cite{2003astro.ph.10890C}. 
The two differ in details:
while \cite{2004MNRAS.350.1107M} use an 
N-body (PM) simulation and are limited to small box sizes, 
\cite{2003astro.ph.10890C} uses a larger $P^3M$ simulation and 
patches onto it sections of a high resolution hydrodynamic simulation. 
A large simulation box is needed to explore fluctuations in
the ionizing background 
from quasars, whose typical separation is tens of megaparsecs. 
High resolution is needed to resolve the \lyaf. 
Here we use a very large TPM simulation with 320$\hmpc$ box size and $1024^3$ 
resolution to obtain both a large box and a high resolution within the same 
simulation \citep{2003ApJS..145....1B}. 
This allows us to explore the background 
fluctuations on large scales and gives correct correlations between 
large and  
small scales. While the simulation does not converge in 
an absolute sense on small scales, it probably converges for the 
purpose of relative 
comparisons between simulation results with and without ionizing background 
fluctuations \citep{2004MNRAS.350.1107M}. 
The cosmological parameters of the simulation are 
$\Omega_m=0.3$, $H_0=70{\rm km/s/Mpc}$ and $\sigma_8=0.93$. 

To generate a realization of the quasar background we 
first select 
the sites of quasar hosts. In \cite{2004MNRAS.350.1107M} positions of 
these were chosen randomly, 
but this ignores the possibility that the \lyaf\ and background fluctuations 
are correlated.  In \cite{2003astro.ph.10890C} 
quasar host sites were chosen based on 
overdensity criteria, because the resolution of the simulation was insufficient 
to resolve galactic size halos. We have actual  
halo catalogs from a halo finder, which are reliable 
down to masses around $10^{11}h^{-1}M_{\sun}$ (the particle 
mass is $3\times 10^{9}h^{-1}M_{\sun}$). 
We use these halos as potential quasar hosts. We assume all quasars 
have the same lifetime $t_{\rm QSO}$, which we will vary 
between $10^7-10^8$ years \citep{2002ApJ...576..653S}. 

We model the quasar luminosity function as a double power law, 
\be
\Phi(L)={\Phi^* \over (L/L_*)^{\beta}+(L/L_*)^{\beta_2}},
\label{lf}
\ee
with the faint end slope $\beta_2=1.58$. 
For the bright end 
we explore two possibilities, $\beta=3.41$ as suggested by 
quasars with $z<2.3$ \citep{2000MNRAS.317.1014B}, and 
$\beta=2.58$ found in high redshift quasars with 
$z>4$ \citep{2001AJ....121...54F}. 
The latter choice is flatter, 
has more bright quasar sources and leads to stronger 
fluctuations. The B band magnitude corresponding to $L_*$ in equation 
\ref{lf} is assumed to be given by $M_*=-22.65-2.5(1.36z-0.27z^2)$, 
while $\Phi_*=0.36 \times 10^{-6} {\rm Mpc}^{-3}{\rm mag}^{-1}$ 
\citep{2003MNRAS.342.1205M}. 

To generate a realization of 
quasars in the simulation at a given redshift
we take the quasar luminosity function and multiply it by the ratio of 
the age of the universe at that redshift, $t_{\rm Hubble}$, 
to the quasar lifetime, $t_{\rm QSO}$. 
This gives us the density of quasar hosts, each of which will be 
a host to an active quasar at any time with a probability of 
$t_{\rm QSO}/t_{\rm Hubble}$. We then assign halo hosts to 
quasar hosts assuming a monotonic relation between the quasar 
luminosity (when active) and halo mass. 
We generate a turn-on time $t_{\rm on}$ for each quasar 
choosing randomly between 0 and $t_{\rm Hubble}$. 
In general we choose quasars down to $M_B=-22$, which is fainter than the 
turnover magnitude $M_*$. 

To compute the background radiation at any given position 
we add up contributions from all quasars. We do this by first computing 
the time of observation $t_{\rm obs}$ 
for a given position along the line of sight. 
We compute the distance $r$ from that position 
to each quasar and convert it to light 
propagation time $t_{\rm prop}$. To count the contribution from a 
given quasar one must have 
$t_{\rm on}+t_{\rm prop}<t_{\rm obs}<t_{\rm on}+t_{\rm prop}+t_{\rm qso}$, 
so that the radiation from the quasar is passing 
through the given position at the time of observation. 
This procedure accounts for the light-cone effects discussed in 
\cite{2003astro.ph.10890C}. 

Each quasar adds a contribution to the radiation field $\Gamma$
proportional to $L~r^{-2}\exp(-r/r_{\rm att})$. The attenuation length 
$r_{\rm att}$ is determined by the overall level of absorption of 
the radiation by neutral hydrogen. The actual value is somewhat 
uncertain and there are various estimates in the literature 
\citep{1993ApJ...415..524F,1996ApJ...461...20H,2003astro.ph.10890C,
2004MNRAS.350.1107M}.  For the low density 
\lyaf\ simulations can be used. In addition there is also 
a significant contribution from high column density 
damped and Lyman-limit systems, which are typically not properly reproduced 
in simulations and their contribution has to be estimated separately. 
Here we will use $r_{\rm att}=1.7\times 10^4 (1+z)^{-3.2}h^{-1}{\rm Mpc}$
\citep{2004MNRAS.350.1107M}, but we will 
also explore the more pessimistic scenario where the attenuation lengths 
are 50\% shorter (at the lower limit of the various estimates in the
literature). In general, shorter attenuation lengths lead to 
larger fluctuations, since nearby sources account for a larger fraction of
the radiation in this case. 

\begin{figure}
\centerline{\psfig{file=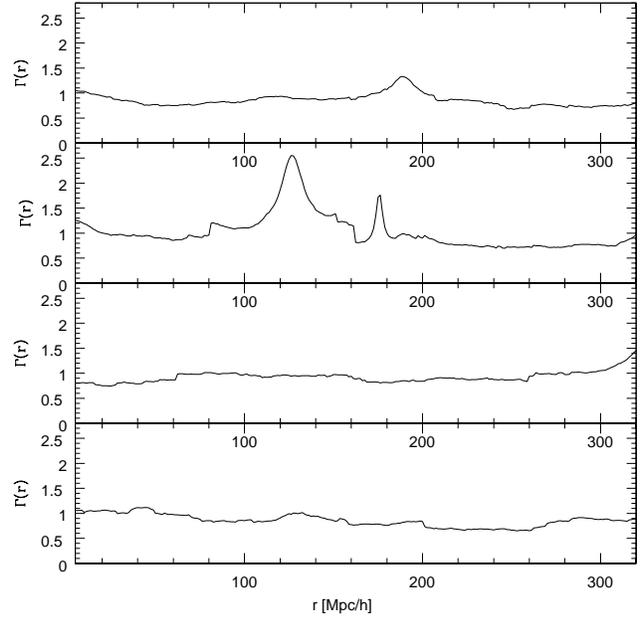,width=3.5in}}
\caption{
Examples of radiation background fluctuations along a line of sight 
assuming all of the background comes from quasars.
This example is for $z=3.29$ with $r_{\rm att}=161h^{-1}{\rm Mpc}$
and quasar lifetime $t=10^8$yr. 
}
\label{uvfluc}
\end{figure}

The attenuation length increases rapidly with decreasing 
redshift and at low redshifts 
it can exceed the simulation box size.  This means that considering only 
the contributions from within the box will underestimate the 
background and so overestimate the fluctuations (since we always 
normalize the background relative to the required level needed to achieve 
a given mean level of absorption). 
We solve this by stacking together several boxes, 
summing over the quasars in all of them. These additional 
boxes are 
identical realizations of the central box and we use periodic 
boundary conditions to patch them together. In total we perform the 
summation over 27 boxes. Note that while the additional boxes are 
identical to the original box, the actual quasars contributing 
to a given position are not repeated, since 
the quasar lifetime is much shorter than the light travel time 
through the box and so no single 
quasar can contribute to a given point more than once. 
Examples of the radiation field along a line of sight 
are shown in figure \ref{uvfluc}, This example is for 
$z=3$, $t_{\rm QSO}=10^8$yr and $r_{\rm att}=161h^{-1}{\rm Mpc}$.
We see that most lines of sight are very smooth with radiation 
density close to the mean, but every now and then the line of 
sight passes very close to an active quasar and the radiation 
density increases. In the most prominent cases one can see the 
light cone effects and times of quasar turn-on and turn-off. 

\begin{figure}
\centerline{\psfig{file=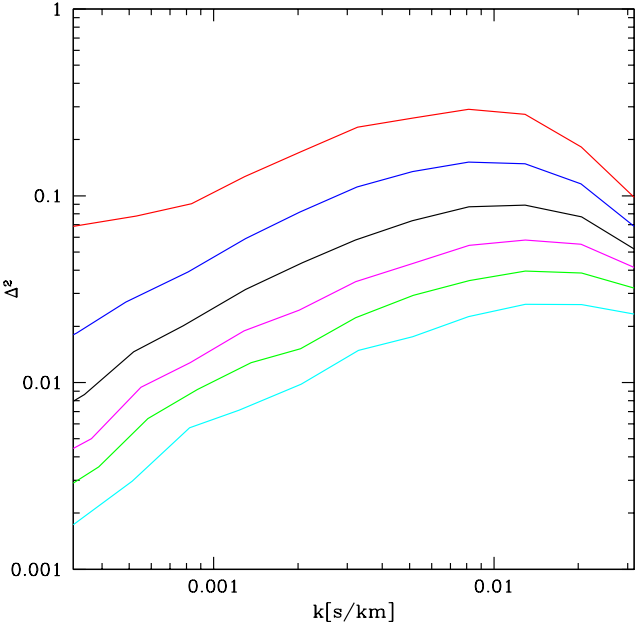,width=3.5in}}
\caption{
Flux power spectrum from the TPM simulation. The redshifts from top to bottom 
are $z=4.58$, 3.88, 3.29, 2.82, 2.40 and 2.05. The spectra agree qualitatively 
well with the observations. 
}
\label{ps}
\end{figure}

To generate \lyaf\ spectra from a pure N-body simulation we 
choose a line of sight and use cloud-in-cell interpolation 
of the nearby particles
onto a one dimensional grid consisting of 1024 cells of the same size
as the original 3-d simulation cells.
This gives us the density and velocity at 
each grid cell. We compute the neutral density/optical depth using the 
ionizing equilibrium relation 
\be
\tau =A {(1+\delta)^{1.7} \over \Gamma},
\ee
where $\delta$ is the dark matter overdensity, $\Gamma$ is the 
intensity of the
radiation background, the factor of 1.7 is approximately valid for
the observed temperature-density relation
(for isothermal gas the slope is 2 and the results are almost identical), 
and $A$ is a constant of proportionality \citep{1997MNRAS.292...27H}. 
For the ionizing background $\Gamma$ we take the local value at that position. 
We then map the neutral density field into redshift space 
by adding the peculiar velocity to the position at each grid cell
and interpolate this field back to the grid. We ignore any additional 
thermal smoothing, since 
the grid cell size is already larger than the thermal broadening scale. 
Finally we compute the flux $F=\exp(-\tau)$. 
We vary the constant $A$ until achieving the desired mean flux decrement,
$\bar{F}$, to match observations (see \cite{2004MNRAS.350.1107M} 
for a recent compilation). 
Figure \ref{ps} shows the resulting power spectra as a function of 
redshift. Comparison to observations reveals good qualitative agreement over the 
whole range of scales and redshifts. When comparing the spectra with 
and without background radiation fluctuations we always match first the mean 
flux for the two cases.

\begin{figure}
\centerline{\psfig{file=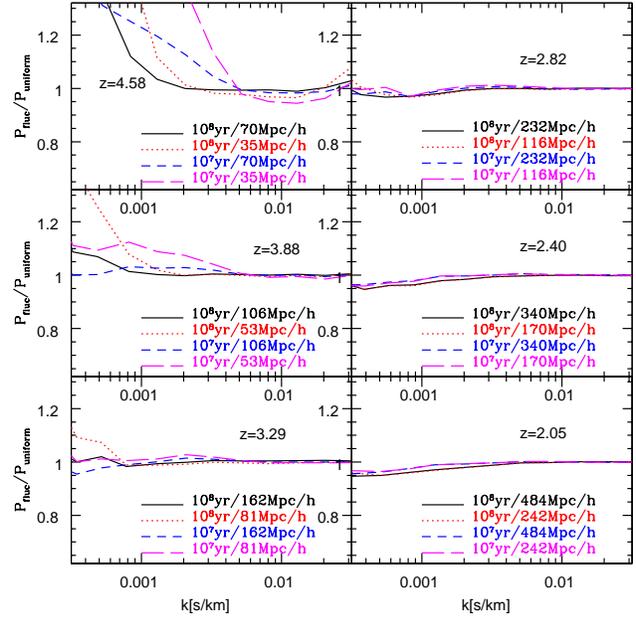,width=3.5in}}
\caption{
Ratio of the power spectra from simulations with ionizing background 
fluctuations versus uniform background, 
for various quasar lifetimes and attenuation lengths.  
This is for bright end 
slope of the quasar luminosity function $\beta=3.41$. 
We assume all of the ionizing background comes from quasars.
}
\label{psrat1}
\end{figure}

\begin{figure}
\centerline{\psfig{file=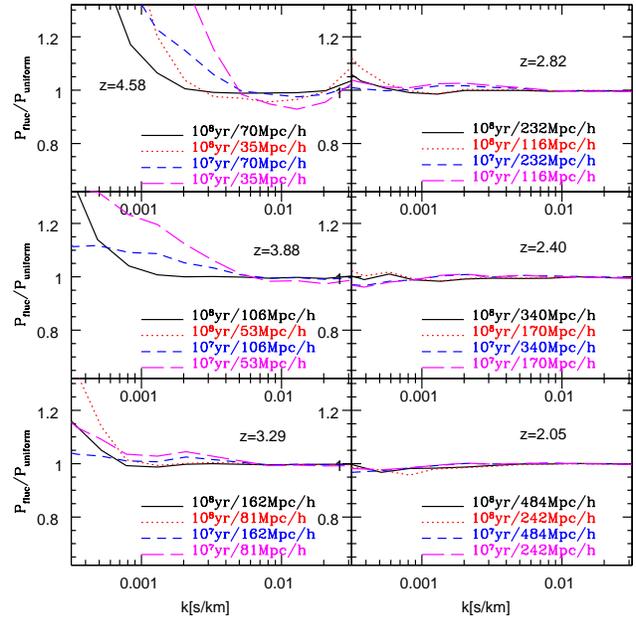,width=3.5in}}
\caption{
Same as figure \ref{psrat1} using bright end slope of the quasar 
luminosity function $\beta=2.58$. 
}
\label{psrat2}
\end{figure}

For each line of sight we compute the flux distribution 
using the uniform background approximation and using the 
actual radiation field at each point. We then compute the 
power spectrum for each case, averaging over thousands lines of sight
in each case. In figures \ref{psrat1} and \ref{psrat2} we present the results
for these ratios for both $\beta=3.41$ and $\beta=2.58$. 
In all cases we show the scenario where the quasar
contribution to the background is 100\%. 
If quasars contribute a fraction $\alpha$, with the rest of the 
background uniform, then the 
effects in figures \ref{psrat1}-\ref{psrat2} are reduced roughly 
by this amount.  

There are several qualitative features in 
figures \ref{psrat1}-\ref{psrat2} worth pointing out. 
First is the fact that at high redshifts where the effect is 
significant the fluctuations in 
ionizing background enhance the power spectrum on large scales. 
A second important feature is the rapid decrease of the importance 
of fluctuations below $z=4$. This is a consequence of the increase
in the attenuation length $r_{\rm att}$, which leads to more 
quasars contributing to a given position and thus reducing the 
fluctuations. 

For the SDSS power spectrum measurements \citep{2004astro.ph..5013M} 
the relevant range is $10^{-3}~{\rm s/km} <k<2\times 10^{-2}~{\rm s/km}$ 
and $2.2<z<4.2$. These observations do not go to sufficiently 
high redshift for these radiation background fluctuation 
effects to be clearly visible in the power spectra. 
There is no evidence for any enhancement of power on large scales at the
highest redshifts ($z=4.2$); note, however, that the statistical errors 
are relatively large here \citep{2004astro.ph..7377M}. 
Some extreme models, like 
$t_{\rm QSO}=10^7$yr, $\beta=2.58$ and $r_{\rm att}=53h^{-1}$Mpc 
produce a 20\% effect at $k=10^{-3}~{\rm s/km}$ and $z=3.88$. 
These models are strongly 
disfavored by the data. Other models, like 
$t_{\rm QSO}=10^8$yr, $\beta=3.4$ and $r_{\rm att}=106h^{-1}$Mpc, 
produce no effect at this redshift and are acceptable even if 
quasars dominate the background radiation field.  
The effects are even smaller at lower redshift. 

Comparing our results to \cite{2003astro.ph.10890C} 
at $z=3$, we note that 
the effect found there is somewhat larger and has the opposite 
sign. It is unclear 
what the source of the difference is, but one possibility is the 
hybrid hydrodynamic-dark matter 
approach used in \cite{2003astro.ph.10890C}, which allows for only 
a qualitative estimate of the effect. 
Another difference in the two treatments is the choice of quasar hosts. 
While we use actual halos, overdensities were used in 
\cite{2003astro.ph.10890C}. Finally there are also some differences in the 
adopted quasar 
luminosity function. While redshift dependence was not explored in 
\cite{2003astro.ph.10890C}, our analysis suggests that a model that 
produces a 20\% effect 
at $k=10^{-3}~{\rm s/km}$ and $z=3$ will lead to a large effect at $z=4$ and 
will be excluded by the absence of any 
such effect in real data \citep{2004astro.ph..5013M}. 

There could be additional sources of ionizing radiation such as
Lyman break galaxies  
\citep{2003ApJ...592..728S} or reradiation of photons from 
the absorbing sources in the forest \citep{1996ApJ...461...20H}. 
It is often assumed that due to their higher density these sources lead 
to uniform radiation field and so their effects on ionizing 
fluctuations need not be considered. However, these sources can still be 
clustered and if the large scale effects are coherent their
contribution could be important. Here we address this by 
choosing all halos with masses between $10^{11}h^{-1}M_{\sun}$ and 
$10^{12}h^{-1}M_{\sun}$ as the hosts of LBGs. Their number density is 
$6 \times 10^{-3}h^{3}{\rm Mpc}^{-3}$, comparable or somewhat larger 
than the density of known LBGs \citep{2003ApJ...592..728S}. We assign a luminosity 
proportional to halo mass and assume long lifetimes, comparable to 
the Hubble time at a given redshift. This model need not be correct, 
since LBGs could be starburst galaxies hosted by small halos 
rather than sitting in the most massive halos at that redshift \citep{2001MNRAS.320..504S}, 
but we lack mass resolution to resolve these small halos and test 
this alternative model. 
It is likely that the starburst model  
leads to more homogeneous radiation background. 

\begin{figure}
\centerline{\psfig{file=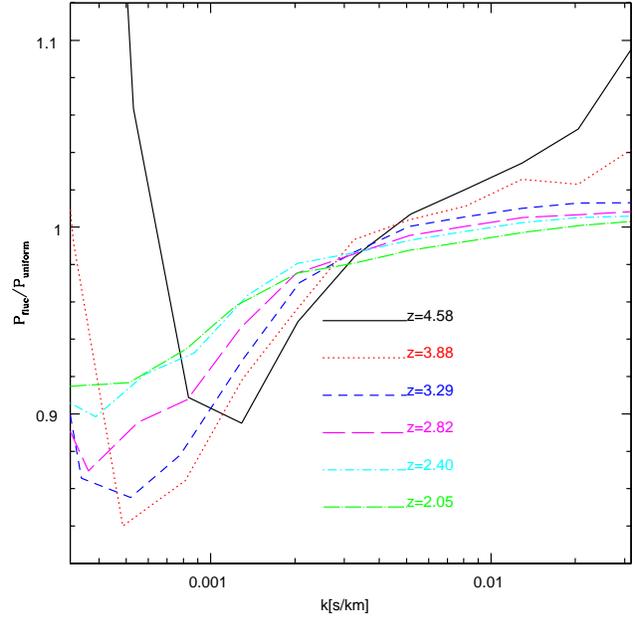,width=3.5in}}
\caption{
Example of the effect on the flux power spectrum 
if the sources of ionizing radiation 
are galaxies with density comparable to that of LBGs. 
We find a suppression of power in this case. 
}
\label{figg}
\end{figure}

The rest of the model is the same as for quasars and in our calculations
we assume all of ionizing background comes from these sources.
In figure \ref{figg} we show the effect on the flux power spectrum
as a function of redshift. We see that the redshift dependence is 
still present, but is weaker than for quasars. The effect is small
except on largest scales, where it can reach 10\% at redshifts of
interest. Interestingly, over the relevant range of scales the effect 
leads to a suppression of power, contrary to the quasar case. This could 
be a consequence of coherence between the galaxy field and \lyaf\, 
which comes into better light for these high density sources, unlike
the case of quasars where shot noise was more important. 

In an actual analysis of the real data \citep{2004astro.ph..7377M} we tried adopting 
the templates of the worst case quasar 
scenario and allowing the fractional quasar contribution to 
vary smoothly with redshift (while the estimates of quasar 
contribution to the total background are uncertain, we do not
expect the relative fraction to vary wildly with redshift 
\citep{2003MNRAS.342.1205M}). 
The data strongly disfavor these models \citep{2004astro.ph..7377M}. 
This could be because the worst case scenario 
quasar model adopted is too extreme, so 
other models in figures \ref{psrat1}-\ref{psrat2}
could still be acceptable as they 
have little or no effect on the flux power spectrum. Alternatively, the quasar 
effect should be combined with the contribution from galaxies, 
which leads to an opposite effect and so the two contributions partially 
cancel out. 

\section{Galactic superwinds}

The effects 
of GSW on the \lyaf\ power spectrum have been explored before 
\citep{2002ApJ...580..634C,2003astro.ph.11209D}, where 
the wind simulation power spectra were compared to those with 
no winds. 
In this paper we 
perform a similar analysis using hydrodynamic 
simulations with and without GSW,
but we also account for the allowed changes in the mean absorption, 
temperature-density relation,
and filtering length. These are part of our standard \lyaf\ analysis 
\citep{2004astro.ph..7377M}
and our goal is to investigate whether there are additional 
effects that go beyond the simple changes in the mean transmitted
flux fraction, $\bF$, mean temperature,
$T_0$, slope of the temperature-density relation, $\gamma-1$,
and filtering length, $k_F$. 

Our simulations (similar, but not the same as those in \S \ref{simhighdense}) 
are based on a fixed grid Eulerian 
cosmological hydrodynamic code with a TVD shock-capturing scheme 
\citep{2004astro.ph..7143C}. They include the effects 
of cooling, heating, star formation, and supernova (SN) feedback. 
The cosmology adopted here is standard $\Lambda$CDM with $\Omega_m=0.29$, 
$\Omega_b=0.047$, $H_0=70{\rm km/s/Mpc}$, and $\sigma_8=0.85$. 
All of the simulations used here have a box size of $11h^{-1}$Mpc in 
comoving units,
$216^3$ dark matter particles, and gas followed on a $432^3$ grid. 
See \cite{2004astro.ph..7143C} for additional simulations and 
resolution tests. 

As described in \cite{2004astro.ph..7143C}, 
there is no attempt to model the subgrid physics in these 
simulations, which would in principle determine
how much of the SN energy is injected into the galaxies and 
how much of it can escape them. 
Instead it is 
assumed the wind energy (and mass) flux is proportional to the star 
formation rate $\dot{M}_*$ and the proportionality constant 
is calibrated 
on observations \citep{2003RMxAC..17...47H,2001ApJ...554..981P}. 
For energy output this implies 
$\dot{E}_{GSW}=e_{GSW}c^2\dot{M}_*$ with $e_{GSW}=3\times 10^{-6}$. 
We assume a similar relation for 
the mass output. These are somewhat uncertain and we want to 
explore worst case scenarios, so we will analyze simulations where 
the outflow energy is increased by 5 to $e_{GSW}=1.5\times 10^{-5}$.

As discussed in \cite{2004astro.ph..7143C}, the wind simulations produce 
significant 
outflows which can heat up the IGM and pollute it with metals at a
level consistent with observations. 
However, they appear to have little effect on the distribution of neutral 
hydrogen. This is because GSW propagate preferentially in the 
directions of low density, so while they heat up the gas in voids, 
they do not affect the structure in higher density regions where  
most of the fluctuations in absorption by neutral hydrogen are 
produced \citep{2002ApJ...578L...5T}. 

\begin{figure}
\centerline{\psfig{file=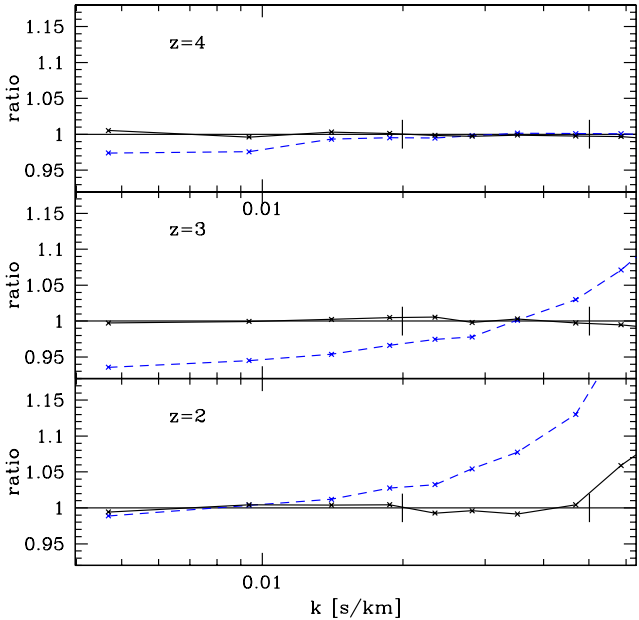,width=3.5in}}
\caption{
Ratio of the power spectra between strong wind simulation and 
no wind simulation for $z=2,3,4$. Dashed line shows direct comparison,
solid line the comparison after adjusting $\bF$, the temperature-density 
relation, and the filtering scale. 
}
\label{fig_wind}
\end{figure}

To quantify these statements we explore explicitly the effect 
of GSW on the \lyaf\ flux power spectrum. This is shown in figure 
\ref{fig_wind} 
for the more extreme value $e_{GSW}=1.5\times 10^{-5}$ 
at 3 different redshifts. 
We do not show the results for the more realistic value 
$e_{GSW}=3\times 10^{-6}$ since 
the effects there are less significant.
While we see that the effects 
for this more extreme case can be
up to 10\% over the range of observational interest, 
especially at $z=3$, the effect is mostly due 
to the fact that the IGM temperature history and density dependence
have changed. 
Without winds, the temperature parameters at $z=$(2, 3, 4) are 
$T_{1.4}=$(21699, 17680, 13478) K and $\gmo=$(0.454, 0.397, 0.497), 
while with maximum winds, they change to
$T_{1.4}=$(19563, 15012, 14455) K and $\gmo=$(0.479, 0.493, 0.477),
where $T_{1.4}$ is the temperature at density 1.4 times the mean.

To explore the effect of changes in the temperature-density relation, 
we recompute the 
power spectra after modifying the temperatures in the simulations to
impose different $T_{1.4}$ and $\gmo$. 
We do this in the following way: First we verify that computing 
the neutral hydrogen density from the gas density using 
the ionizing equilibrium assumption leads 
to identical results as the simulation outputs themselves. 
Then we compute $T_{1.4}$ and $\gmo$ for a simulation 
using a least absolute
deviation fit \citep{1992nrca.book.....P} to all the points
in the simulation with $1<\rho/\bar{\rho}<2$.
This fit effectively uses the median rather 
than the mean temperature, which is desirable because 
the mean temperature values are heavily biased by
rare regions with extremely high temperatures in 
simulations with GSW. 
For each gas cell we compute from its density the 
expected temperature for the 
original temperature-density relation in the simulation 
and for a modified temperature-density relation. We take the
ratio between the two and multiply the actual value of the 
temperature in the cell by it, therefore preserving the scatter in the 
temperature-density relation present in the hydrodynamic simulation. We 
then use the new temperatures to compute the neutral density using 
the ionizing equilibrium equation, and finally recompute the flux power 
spectrum. 

To see how much of the difference between simulations with and
without winds is accounted for by simple temperature-density relation
changes, we 
recompute the power spectra in the simulations with GSW for a 
variety of modified temperature-density 
relations until we find one that matches best the power spectrum 
of the simulation without GSW. We find that the required changes
in $T_{1.4}$ and $\gmo$ are
in similar to the changes in the median-fitted 
values, but 
we did not restrict ourself to these in our analysis since it 
is not clear that they are the appropriate averages. 
For the purpose of the power spectrum analysis,
finding the values that minimize the power spectrum 
differences is most relevant, 
since in our power spectrum analysis we always marginalize 
over $T_{1.4}$ and $\gmo$. 

While we find good agreement between $P_F$ in wind and no-wind
simulations using this procedure where we only
modify $T_{1.4}$ and $\gmo$, 
we can improve it further by noting that a
change in the temperatures also changes the filtering 
length $k_F$. This filtering length depends on the full temperature history
of the IGM from reionization onwards. We do not try to 
make analytic predictions of what the filtering length should 
be given the 
temperature history of the IGM due to the difficulties in interpreting 
the scatter in temperature at a given density. Instead we follow 
the procedure in \cite{2004astro.ph..7377M}, using hydro-PM simulations 
\citep{1998MNRAS.296...44G} to find the effect of changing
the reionization redshift.  
We apply the change that minimizes the 
residual effect of the winds when combined with changes in
$T_{1.4}$ and $\gmo$.  Finally, we also allow small adjustments in the
mean absorption level, $\bF$, which must be marginalized over in
any cosmological fit.
The final result is shown in figure \ref{fig_wind}, where
we see that 
using the freedom of varying $\bF$, $T_{1.4}$, $\gmo$, and $k_F$ cancels 
completely the effect of GSW in the simulations over the region of 
observational interest.  

The parameter variations we used to produce this agreement are not extreme. 
Beyond the change in median temperature-density
relation estimated directly from the simulations, the modifications 
needed are as follows:  We modified
the filtering ($k_F$) effect by 20\% of the difference between two 
cases: the case when reionization heats the gas to 25000 K at $z=7$ 
and the case when reionization heats the gas to 50000 K at $z=17$
(we present the change this way because we interpolated between these
two models instead of actually running intermediate 
simulations \citep{2004astro.ph..7377M}).  
Note that we do not adjust $k_F$ independently
at each redshift -- the early-time thermal history is 
modified and this effects each of the redshifts we show in a 
self-consistent way.
At $z=2$, we make the additional small modifications  
$\Delta \bF=-0.001$, $\Delta T_{1.4}=-800$ K, and 
$\Delta \gmo = -0.03$;
at $z=3$, we make the modifications  
$\Delta \bF=+0.006$, $\Delta T_{1.4}=-1000$ K, and 
$\Delta \gmo = -0.01$; and finally, 
at $z=4$, we make the changes  
$\Delta \bF=+0.0065$, $\Delta T_{1.4}=+1000$ K, and 
$\Delta \gmo = 0.05$.  All of these changes are small
relative to the uncertainty in our measurements of 
these parameters \citep{2004astro.ph..7377M}.

While the simulation box is too small to be able
to probe larger scales, the flatness 
of the final result on large scales suggests that 
these effects are only important on small scales and that these 
conclusions are unlikely to be modified if larger simulation boxes were used.
Note that we find no effect both in the simulation with the expected GSW
amplitude, as well as in the one where it was increased by a factor of 5. 
We only show the worse case scenario in figure \ref{fig_wind}. 

While these results are comforting, they should not be viewed as 
a conclusive proof that GSW have no effect on \lyaf\ flux power 
spectrum. For one thing, it is not clear that our simulations  
fare any better than 
other simulations in explaining the flux enhancement observed in
lines of sight very close to LBGs \citep{2003ApJ...584...45A}.
While no realistic simulations
to date have been able to reproduce this result entirely, it is
possible that it is a consequence of a statistical fluctuation, 
hosts of LBGs being less massive than usually assumed, 
galaxy proximity effect,
selection effects, or any combination of these. Moreover, 
the fact that \cite{2003astro.ph.11209D} come close to explaining these 
observations without affecting the flux power spectrum is 
comforting. 
However, it is also possible that we are 
still missing some essential physics and that the winds are much more 
destructive of the IGM, although there are additional 
constraints on this like the flux 
probability distribution statistics \citep{2000ApJ...543....1M}. 
Even in such a case it is not clear  
how much of the effect is 
degenerate with the parameters we already marginalized over, such as 
$\bF$, $T_{1.4}$, $\gmo$, and $k_F$. It is clear that more work is needed to 
exclude all possible sources of contamination of the \lyaf\ 
flux power spectrum, even if the current modeling suggests that GSW
are not problematic.

\section{Conclusions} 

In this paper we have analyzed physical effects on the power spectrum 
of the \lyaf.  We focus on effects that would not otherwise be 
part of the standard 
simulation-based analysis \citep{2004astro.ph..7377M}. 
In section 2 we analyze the contribution from high density systems,
especially those with damping wings, which 
are not simulated properly in the current generation of simulations. 
We show that their inclusion adds power on large 
scales and accounting for this leads to an increase in  the deduced
slope of primordial fluctuations \citep{2004astro.ph..7377M}.  
The amplitude of the 
effect can be estimated both from direct counting of these high  
density systems as well as using the templates in 
the power spectrum analysis itself, taking
advantage of their specific form as a function of scale and redshift. At the 
moment the uncertainties in the former make the 
latter approach competitive or even better \citep{2004astro.ph..7377M}, 
suggesting that a better determination of 
the high density systems as a function of column density and redshift would 
be useful for accurate determination of this effect.
But even within the current uncertainties it is clear that the effect is 
appreciable compared to the size of the statistical error given by the new 
data. 

In section 3 we analyze the effects of fluctuations in the ionizing 
background. We focus on quasars, which are believed to be the dominant source
of ionizing background at these redshifts. These fluctuations lead to an 
enhancement of power on large scales which
rapidly increases with redshift, because the photon attenuation length 
is significantly shorter at higher redshifts. This allows us to develop 
templates that can be used to identify the effect in the real data. 
The SDSS data show no evidence of this effect \citep{2004astro.ph..7377M}. We also 
investigated a simple example that could correspond to galactic
sources of radiation, matching the density of LBG galaxies. 
Contrary to naive expectations we do not find that 
this more dense population of sources 
leads to a more homogeneous ionizing background, likely
as a result of clustering of sources. The coherent nature of 
these sources and \lyaf\ leads to a suppression of power on large scales. 
The effect is relatively small on scales probed by SDSS 
and is somewhat redshift dependent, which allows it to be constrained 
from the data. 

We should caution that the current level of modeling remains 
rather simplistic and there are many possible scenarios we did 
not explore, so we cannot conclusively rule out that fluctuations 
in ionizing background
are unimportant for the flux 
power spectrum statistics. 
To put things into perspective we note that the effect 
of damped systems, which is 20\% at $k \sim 10^{-3}{\rm s/km}$ (figure \ref{mockHD}),
results in a 0.06 change in the 
slope of the power spectrum, which is slightly more than 1-$\sigma$
\citep{2004astro.ph..7377M}. In comparison to this the effects 
from ionizing fluctuations are below 10\% at $k \sim 10^{-3}{\rm s/km}$.
Thus these effects are unlikely to change the estimate of the slope  
by more than 1-$\sigma$ given the current uncertainties. 
In addition, they have stronger redshift 
dependence than damped systems, 
which makes it possible to identify and remove them. 
It is however important to understand them better if we want to 
extract all of the information present in the SDSS data, as the 
current constraints on the slope  
are limited by theoretical modeling uncertainties and could be 
improved further. 

In contrast to the effects of damped systems and ionizing background 
fluctuations, which mostly change the power spectrum on large scales, 
galactic superwinds are predominantly affecting small scales, where 
several other parameters are also important. 
In section 4 we have shown that galactic superwinds
used in our simulations
do not affect the \lyaf\ flux 
power spectrum statistics after marginalization over the 
mean absorption level, temperature-density relation of the 
IGM, and its filtering scale. 
This is true even though the same GSW can heat up 
a significant fraction of the volume 
and pollute the IGM with metals at a level consistent with observations
\citep{2004astro.ph..7143C}.
This again is not a conclusive proof that winds are unimportant and 
there could still be physics missing in these simulations, but a 
definitive answer can only come by investigating it in 
simulations which are sufficiently realistic and are 
able to satisfy all of the observational 
constraints. 

In summary, current models suggest that various physical effects 
analyzed in this paper, which 
are not part of the current simulations, are not destructive for 
the power spectrum statistics, even if they may weaken its predictive power. 
This conclusion is preliminary and more work is 
needed to investigate these and 
other effects before their 
impact is fully understood and corrected for in the analysis. 
One of the effects we did not investigate here are fluctuations in 
the IGM temperature, which could be caused by patchy 
helium reionization at these redshifts \citep{2002ApJ...564..153Z}. 
This is unlikely to be a major effect, since 
temperature has only a minor effect 
on the flux power spectrum statistics on large scales probed by SDSS, 
but this should be verified with explicit calculations. 
Progress on these issues will come from both theoretical and 
observational directions. 
Better theoretical modeling of the effects discussed in this paper,
with simulations 
that include more realistic physics, will allow a better 
assessment of their impact. Improvements in observational tests of 
the \lyaf\ will be equally important.  
A few examples of these that go beyond the flux power spectrum 
are 
of abundances of damped systems, correlations between \lyaf\ and
quasars or galaxies and higher order correlations of \lyaf\, 
such as the bispectrum analysis or one-point distribution of the flux.  
All of these 
will allow to better constrain the contribution of these effects
to the flux power spectrum and determine 
their effect on the cosmological constraints 
derived from \lyaf\ .

We thank Zheng Zheng for providing the column density distribution
templates, and Jason Prochaska for the tables of DLA data.
We thank Joop Schaye for helpful discussions.
Some of the computations used facilities at Princeton provided in part 
by NSF grant AST-0216105, and some computations were performed at NCSA.
US is supported by a fellowship from the
David and Lucile Packard Foundation,
NASA grants NAG5-1993, NASA NAG5-11489 and NSF grant CAREER-0132953.
RC acknowledges grants AST-0206299 and NAG5-13381.

\bibliography{apjmnemonic,cosmo,cosmo_preprints}   

\begin{thebibliography}{}

\bibitem[\protect\citeauthoryear{{Adelberger} et~al.}{{Adelberger}
  et~al.}{2003}]{2003ApJ...584...45A}
{Adelberger} K.~L., {Steidel} C.~C., {Shapley} A.~E.,  {Pettini} M., 2003,
  \apj, 584, 45

\bibitem[\protect\citeauthoryear{{Aracil} et~al.}{{Aracil}
  et~al.}{2004}]{2004A&A...419..811A}
{Aracil} B., {Petitjean} P., {Pichon} C.,  {Bergeron} J., 2004, \aap, 419, 811

\bibitem[\protect\citeauthoryear{{Bode} \& {Ostriker}}{{Bode} \&
  {Ostriker}}{2003}]{2003ApJS..145....1B}
{Bode} P.,  {Ostriker} J.~P., 2003, \apjs, 145, 1

\bibitem[\protect\citeauthoryear{{Boyle} et~al.}{{Boyle}
  et~al.}{2000}]{2000MNRAS.317.1014B}
{Boyle} B.~J., {Shanks} T., {Croom} S.~M., {Smith} R.~J., {Miller} L.,
  {Loaring} N.,  {Heymans} C., 2000, \mnras, 317, 1014

\bibitem[\protect\citeauthoryear{{Bruscoli} et~al.}{{Bruscoli}
  et~al.}{2003}]{2003MNRAS.343L..41B}
{Bruscoli} M., {Ferrara} A., {Marri} S., {Schneider} R., {Maselli} A.,
  {Rollinde} E.,  {Aracil} B., 2003, \mnras, 343, L41

\bibitem[\protect\citeauthoryear{{Cen}, {Nagamine}, \& {Ostriker}}{{Cen}
  et~al.}{2004}]{2004astro.ph..7143C}
{Cen} R., {Nagamine} K.,  {Ostriker} J.~P., 2004, ArXiv Astrophysics e-prints,
  astro-ph/0407143

\bibitem[\protect\citeauthoryear{{Cen} et~al.}{{Cen}
  et~al.}{2003}]{2003ApJ...598..741C}
{Cen} R., {Ostriker} J.~P., {Prochaska} J.~X.,  {Wolfe} A.~M., 2003, \apj, 598,
  741

\bibitem[\protect\citeauthoryear{{Croft}}{{Croft}}{2003}]{2003astro.ph.10890C}
{Croft} R.~A.~C., 2003, ArXiv Astrophysics e-prints, astro-ph/0310890

\bibitem[\protect\citeauthoryear{{Croft} et~al.}{{Croft}
  et~al.}{2002a}]{2002ApJ...580..634C}
{Croft} R.~A.~C., {Hernquist} L., {Springel} V., {Westover} M.,  {White} M.,
  2002a, \apj, 580, 634

\bibitem[\protect\citeauthoryear{{Croft} et~al.}{{Croft}
  et~al.}{2002b}]{2002ApJ...581...20C}
{Croft} R.~A.~C., {Weinberg} D.~H., {Bolte} M., {Burles} S., {Hernquist} L.,
  {Katz} N., {Kirkman} D.,  {Tytler} D., 2002b, \apj, 581, 20

\bibitem[\protect\citeauthoryear{{Croft} et~al.}{{Croft}
  et~al.}{1998}]{1998ApJ...495...44C}
{Croft} R.~A.~C., {Weinberg} D.~H., {Katz} N.,  {Hernquist} L., 1998, \apj,
  495, 44

\bibitem[\protect\citeauthoryear{{Croft} et~al.}{{Croft}
  et~al.}{1999}]{1999ApJ...520....1C}
{Croft} R.~A.~C., {Weinberg} D.~H., {Pettini} M., {Hernquist} L.,  {Katz} N.,
  1999, \apj, 520, 1

\bibitem[\protect\citeauthoryear{{Desjacques} et~al.}{{Desjacques}
  et~al.}{2003}]{2003astro.ph.11209D}
{Desjacques} V., {Nusser} A., {Haehnelt} M.~G.,  {Stoehr} F., 2003, ArXiv
  Astrophysics e-prints, astro-ph/0311209

\bibitem[\protect\citeauthoryear{{Fan} et~al.}{{Fan}
  et~al.}{2001}]{2001AJ....121...54F}
{Fan} X. et~al., 2001, \aj, 121, 54

\bibitem[\protect\citeauthoryear{{Fardal} \& {Shull}}{{Fardal} \&
  {Shull}}{1993}]{1993ApJ...415..524F}
{Fardal} M.~A.,  {Shull} J.~M., 1993, \apj, 415, 524

\bibitem[\protect\citeauthoryear{{Gardner} et~al.}{{Gardner}
  et~al.}{2001}]{2001ApJ...559..131G}
{Gardner} J.~P., {Katz} N., {Hernquist} L.,  {Weinberg} D.~H., 2001, \apj, 559,
  131

\bibitem[\protect\citeauthoryear{{Gnedin} \& {Hamilton}}{{Gnedin} \&
  {Hamilton}}{2002}]{2002MNRAS.334..107G}
{Gnedin} N.~Y.,  {Hamilton} A.~J.~S., 2002, \mnras, 334, 107

\bibitem[\protect\citeauthoryear{{Gnedin} \& {Hui}}{{Gnedin} \&
  {Hui}}{1998}]{1998MNRAS.296...44G}
{Gnedin} N.~Y.,  {Hui} L., 1998, \mnras, 296, 44

\bibitem[\protect\citeauthoryear{{Haardt} \& {Madau}}{{Haardt} \&
  {Madau}}{1996}]{1996ApJ...461...20H}
{Haardt} F.,  {Madau} P., 1996, \apj, 461, 20

\bibitem[\protect\citeauthoryear{{Heckman}}{{Heckman}}{2003}]{2003RMxAC..17...%
47H}
{Heckman} T.~M., 2003, in Revista Mexicana de Astronomia y Astrofisica
  Conference Series

\bibitem[\protect\citeauthoryear{{Hui} \& {Gnedin}}{{Hui} \&
  {Gnedin}}{1997}]{1997MNRAS.292...27H}
{Hui} L.,  {Gnedin} N.~Y., 1997, \mnras, 292, 27

\bibitem[\protect\citeauthoryear{{Kim} et~al.}{{Kim}
  et~al.}{2003}]{2003astro.ph..8103K}
{Kim} T.~., {Viel} M., {Haehnelt} M.~G., {Carswell} R.~F.,  {Cristiani} S.,
  2003

\bibitem[\protect\citeauthoryear{{Kollmeier} et~al.}{{Kollmeier}
  et~al.}{2003}]{2003ApJ...594...75K}
{Kollmeier} J.~A., {Weinberg} D.~H., {Dav{\' e}} R.,  {Katz} N., 2003, \apj,
  594, 75

\bibitem[\protect\citeauthoryear{{Maselli} et~al.}{{Maselli}
  et~al.}{2004}]{2004MNRAS.350L..21M}
{Maselli} A., {Ferrara} A., {Bruscoli} M., {Marri} S.,  {Schneider} R., 2004,
  \mnras, 350, L21

\bibitem[\protect\citeauthoryear{{McDonald}, {Miralda-Escud{\' e}}, \&
  {Cen}}{{McDonald} et~al.}{2002}]{2002ApJ...580...42M}
{McDonald} P., {Miralda-Escud{\' e}} J.,  {Cen} R., 2002, \apj, 580, 42

\bibitem[\protect\citeauthoryear{{McDonald} et~al.}{{McDonald}
  et~al.}{2000}]{2000ApJ...543....1M}
{McDonald} P., {Miralda-Escud{\' e}} J., {Rauch} M., {Sargent} W.~L.~W.,
  {Barlow} T.~A., {Cen} R.,  {Ostriker} J.~P., 2000, \apj, 543, 1

\bibitem[\protect\citeauthoryear{{McDonald} et~al.}{{McDonald}
  et~al.}{2004a}]{2004astro.ph..5013M}
{McDonald} P. et~al., 2004a, ArXiv Astrophysics e-prints, astro-ph/0405013

\bibitem[\protect\citeauthoryear{{McDonald} et~al.}{{McDonald}
  et~al.}{2004b}]{2004astro.ph..7377M}
{McDonald} P., {Seljak} U., {Cen} R.,  {the SDSS Collaboration} , 2004b,
  astro-ph/0407377

\bibitem[\protect\citeauthoryear{{Meiksin} \& {White}}{{Meiksin} \&
  {White}}{2001}]{2001MNRAS.324..141M}
{Meiksin} A.,  {White} M., 2001, \mnras, 324, 141

\bibitem[\protect\citeauthoryear{{Meiksin} \& {White}}{{Meiksin} \&
  {White}}{2003}]{2003MNRAS.342.1205M}
{Meiksin} A.,  {White} M., 2003, \mnras, 342, 1205

\bibitem[\protect\citeauthoryear{{Meiksin} \& {White}}{{Meiksin} \&
  {White}}{2004}]{2004MNRAS.350.1107M}
{Meiksin} A.,  {White} M., 2004, \mnras, 350, 1107

\bibitem[\protect\citeauthoryear{{Miralda-Escude} et~al.}{{Miralda-Escude}
  et~al.}{1996}]{1996ApJ...471..582M}
{Miralda-Escude} J., {Cen} R., {Ostriker} J.~P.,  {Rauch} M., 1996, \apj, 471,
  582

\bibitem[\protect\citeauthoryear{{Nagamine}, {Springel}, \&
  {Hernquist}}{{Nagamine} et~al.}{2004}]{2004MNRAS.348..421N}
{Nagamine} K., {Springel} V.,  {Hernquist} L., 2004, \mnras, 348, 421

\bibitem[\protect\citeauthoryear{{P{\' e}roux} et~al.}{{P{\' e}roux}
  et~al.}{2003a}]{2003MNRAS.345..480P}
{P{\' e}roux} C., {Dessauges-Zavadsky} M., {D'Odorico} S., {Kim} T.,  {McMahon}
  R.~G., 2003a, \mnras, 345, 480

\bibitem[\protect\citeauthoryear{{P{\' e}roux} et~al.}{{P{\' e}roux}
  et~al.}{2003b}]{2003MNRAS.346.1103P}
{P{\' e}roux} C., {McMahon} R.~G., {Storrie-Lombardi} L.~J.,  {Irwin} M.~J.,
  2003b, \mnras, 346, 1103

\bibitem[\protect\citeauthoryear{{Pettini} et~al.}{{Pettini}
  et~al.}{2001}]{2001ApJ...554..981P}
{Pettini} M., {Shapley} A.~E., {Steidel} C.~C., {Cuby} J., {Dickinson} M.,
  {Moorwood} A.~F.~M., {Adelberger} K.~L.,  {Giavalisco} M., 2001, \apj, 554,
  981

\bibitem[\protect\citeauthoryear{{Press} et~al.}{{Press}
  et~al.}{1992}]{1992nrca.book.....P}
{Press} W.~H., {Teukolsky} S.~A., {Vetterling} W.~T.,  {Flannery} B.~P., 1992,
  {Numerical recipes in C. The art of scientific computing}.
\newblock Cambridge: University Press, |c1992, 2nd ed.

\bibitem[\protect\citeauthoryear{{Prochaska} \& {Herbert-Fort}}{{Prochaska} \&
  {Herbert-Fort}}{2004}]{2004astro.ph..3391P}
{Prochaska} J.~X.,  {Herbert-Fort} S., 2004, ArXiv Astrophysics e-prints,
  astro-ph/0403391

\bibitem[\protect\citeauthoryear{{Schaye}}{{Schaye}}{2001}]{2001ApJ...559..507%
S}
{Schaye} J., 2001, \apj, 559, 507

\bibitem[\protect\citeauthoryear{{Schaye} et~al.}{{Schaye}
  et~al.}{2003}]{2003ApJ...596..768S}
{Schaye} J., {Aguirre} A., {Kim} T., {Theuns} T., {Rauch} M.,  {Sargent}
  W.~L.~W., 2003, \apj, 596, 768

\bibitem[\protect\citeauthoryear{{Smith}, {Cohen}, \& {Bradley}}{{Smith}
  et~al.}{1986}]{1986ApJ...310..583S}
{Smith} H.~E., {Cohen} R.~D.,  {Bradley} S.~E., 1986, \apj, 310, 583

\bibitem[\protect\citeauthoryear{{Somerville}, {Primack}, \&
  {Faber}}{{Somerville} et~al.}{2001}]{2001MNRAS.320..504S}
{Somerville} R.~S., {Primack} J.~R.,  {Faber} S.~M., 2001, \mnras, 320, 504

\bibitem[\protect\citeauthoryear{{Songaila} \& {Cowie}}{{Songaila} \&
  {Cowie}}{1996}]{1996AJ....112..335S}
{Songaila} A.,  {Cowie} L.~L., 1996, \aj, 112, 335

\bibitem[\protect\citeauthoryear{{Steidel} et~al.}{{Steidel}
  et~al.}{2003}]{2003ApJ...592..728S}
{Steidel} C.~C., {Adelberger} K.~L., {Shapley} A.~E., {Pettini} M., {Dickinson}
  M.,  {Giavalisco} M., 2003, \apj, 592, 728

\bibitem[\protect\citeauthoryear{{Steidel} et~al.}{{Steidel}
  et~al.}{2002}]{2002ApJ...576..653S}
{Steidel} C.~C., {Hunt} M.~P., {Shapley} A.~E., {Adelberger} K.~L., {Pettini}
  M., {Dickinson} M.,  {Giavalisco} M., 2002, \apj, 576, 653

\bibitem[\protect\citeauthoryear{{Theuns} et~al.}{{Theuns}
  et~al.}{2002}]{2002ApJ...578L...5T}
{Theuns} T., {Viel} M., {Kay} S., {Schaye} J., {Carswell} R.~F.,  {Tzanavaris}
  P., 2002, \apjl, 578, L5

\bibitem[\protect\citeauthoryear{{Viel} et~al.}{{Viel}
  et~al.}{2004}]{2004MNRAS.349L..33V}
{Viel} M., {Haehnelt} M.~G., {Carswell} R.~F.,  {Kim} T.-S., 2004, \mnras, 349,
  L33

\bibitem[\protect\citeauthoryear{{Wolfe} et~al.}{{Wolfe}
  et~al.}{1986}]{1986ApJS...61..249W}
{Wolfe} A.~M., {Turnshek} D.~A., {Smith} H.~E.,  {Cohen} R.~D., 1986, \apjs,
  61, 249

\bibitem[\protect\citeauthoryear{{Zaldarriaga}}{{Zaldarriaga}}{2002}]{2002ApJ.%
..564..153Z}
{Zaldarriaga} M., 2002, \apj, 564, 153

\bibitem[\protect\citeauthoryear{{Zheng} \& {Miralda-Escud{\' e}}}{{Zheng} \&
  {Miralda-Escud{\' e}}}{2002}]{2002ApJ...568L..71Z}
{Zheng} Z.,  {Miralda-Escud{\' e}} J., 2002, \apjl, 568, L71

\bibitem[\protect\citeauthoryear{{Zuo}}{{Zuo}}{1992}]{1992MNRAS.258...36Z}
{Zuo} L., 1992, \mnras, 258, 36

\end{thebibliography}
\bibliographystyle{mnras}

\end{document}